\documentclass[
 reprint,
superscriptaddress,
groupedaddress,
amsmath,amssymb,
aps,
pra,
onecolumn,
showkeys
]{revtex4-2}
\usepackage{graphicx}
\usepackage{dcolumn}
\usepackage{xcolor}
\usepackage{bm}
\usepackage[title]{appendix}
\usepackage{lipsum}
\usepackage{hyperref}
\usepackage{cleveref} 
\begin{document}


\title{Incompressible Extended Magnetohydrodynamics Waves: Implications of Electron Inertia}

\author{Noura E. Shourba}
 \affiliation{Theoretical Physics Group, Physics Department, Faculty of Science, Mansoura University, Mansoura, 35516 Egypt.}
\author{Abeer A. Mahmoud}%
\affiliation{Theoretical Physics Group, Physics Department, Faculty of Science, Mansoura University, Mansoura, 35516 Egypt.}%


\author{Hamdi M. Abdelhamid}
\email{hamdi_mtprg@mans.edu.eg}
\affiliation{Theoretical Physics Group, Physics Department, Faculty of Science, Mansoura University, Mansoura, 35516 Egypt.}%
\affiliation{Physics Department, Faculty of Science, New Mansoura University, New Mansoura City, Egypt.}%


\date{\today}

\begin{abstract}
 This paper explores plasma wave modes using the extended magnetohydrodynamics (XMHD) model, incorporating Hall drift and electron inertia effects. We utilize the geometric optics ansatz to study perturbed quantities, with a focus on incompressible systems. Our research concludes with the derivation of the dispersion relation for incompressible XMHD and the associated eigenvector solutions, offering new perspectives on plasma wave behavior under these extended scenarios. The dispersion relation shows distinct ion cyclotron and whistler wave branches, with characteristic saturation at the ion and electron gyrofrequencies, respectively. Comparisons between Hall MHD and XMHD demonstrate that XMHD provides a more accurate representation of plasma dynamics, especially at higher wave numbers, bridging the gap between simplified models and comprehensive two-fluid descriptions and smoothing out singularities present in Hall MHD solutions and capturing more physics of the full two-fluid model. 
\end{abstract}

\maketitle


\section{\label{Intro}Introduction}

The complexity inherent in plasma physics, particularly in the context of extended magnetohydrodynamics (XMHD) \cite{Kimura2014, Abdelhamid2015, Keramidas1, Lingam2015b, Hirota2021}, arises from a scale hierarchy influenced by small-scale effects such as Hall drift and electron inertia. These scale interactions are critical in understanding various plasma processes, including magnetic reconnection \cite{Grasso2017, Nahuelreconnection1, Nahuelreconnection2}, turbulence \cite{Abdelhamid2016, Lingam2016Turb, Nahuelturb, Andres}, dynamos and outflows \cite{Lingam2015Outflow, Mahajan_2005}, which are driven by the interplay of different scales. Such processes emphasize the significant impact of microscopic phenomena on the macroscopic evolution of structures within plasma systems. Recent investigations have shown that electron-mediated processes have a substantial impact on plasma behavior at all scales \cite{Verscharen2022}. Observations of astrophysical phenomena such as turbulence \cite{Bandyopadhyay2018, Chasapis2017, Hadid2015, Hadid2018, Ergun2018, Stawarz2019}, magnetic reconnection \cite{Lu2020, Ji2022, Wang2023, Webster2018, Yan2022, Torbert2018, Eriksson2016, Zweibel2016}, and atmospheric escape \cite{Gronoff2020, Brain2016} have shown the existence of electron-scale dynamics that lead to the formation of large structures. Furthermore, laboratory experiments in a magnetized plasma device have shown the fundamental importance of electron physics in understanding plasma heating and shaping magnetic field configuration \cite{Rigby2018, Mattoo2012}. These investigations highlight the significance of electron-scale interactions in both natural and controlled environments, thereby closing the gap between small-scale physics and large-scale cosmic phenomena.
\\

Ideal magnetohydrodynamics (MHD), considered the fundamental single-fluid model for plasma behavior in strong magnetic fields, combines fluid hydrodynamics with classical electrodynamics (Maxwell equations) \cite{freidberg1987ideal}. Although it is widely used to describe plasma equilibrium and stability in various scenarios (see, e.g., Refs.\cite{goedbloed2019magnetohydrodynamics, goedbloed2004principles}), ideal MHD often fails to account for phenomena arising from different scale hierarchies, especially those influenced by ion and electron inertial lengths. For example, in high-temperature (collisionless) plasma, electron inertia can cause topological changes in magnetic field lines at small scales, a phenomenon that ideal MHD overlooks \cite{Grasso2017,Andres,Abdelhamid20162,Abdelhamid2017}.
\\

Extended MHD (XMHD), which includes effects like Hall drift and electron inertia, introduces dispersion into the wave modes, extending beyond the capabilities of ideal MHD. In this framework, Hall MHD, characterized by the inclusion of the Hall current in the induction equation and the assumption of massless electron ($m_e \rightarrow 0$), diverges from ideal MHD by introducing a short length scale - the ion skin depth. This development allows for the propagation of diverse wave types, with varying phase velocities; For a detailed analysis of HMHD waves, refer to \cite{Hameiri2005}. XMHD further incorporates both electron and ion skin depths, leading to new wave modes such as kinetic and inertial Alfvén waves \cite{cramer2011physics, Louarn1994, roytershteyn2019numerical}.
\\

This study aims to clarify the effects of electron inertia on the dispersion relations within XMHD models. Typically, in the standard ideal MHD framework, arbitrary waveforms are present due to its lack of dispersion \cite{Yoshida2012, cramer2011physics}. However, adding the Hall term introduces notable dispersion, dividing the waves into three separate modes \cite{Hameiri2005, Kawazura2017R}. Our research concentrates on the combined effects of electron inertia and the Hall term, employing the geometric optics approach to understand these complex interactions \cite{Bernstein1975, Hameiri2005}. For a comprehensive understanding of geometric optics technique, we direct readers to the following books \cite{KravtsovOrlov1990, Whitham1999, Swanson_2003}. Given the intricate nature of the XMHD model, we analyze the incompressible limit in conjunction with inhomogeneous equilibrium fields. This targeted method enables us to thoroughly investigate the changes in wave behavior resulting from the inclusion of electron inertia.
 \\

The structure of our paper is organized as follows: Section \ref{XMHD} introduces the mathematical preliminaries of the incompressible XMHD governing equations. Section \ref{waves} focuses on the derivation of incompressible XMHD dispersion relations and the associated eigenvectors, utilizing the geometric optics ansatz. This section also discusses various limiting cases. The paper ends in Section \ref{conclusion}, where we summarize and conclude our findings.

\section{\label{XMHD} Incompressible XMHD: The mathematical preliminaries}

The comprehensive derivation of the XMHD model has been  comprehensively presented in \cite{Abdelhamid2015}, with additional elucidation provided in \cite{Kimura2014}. It is crucial to understand that the governing equations of XMHD are normalized in Alfvénic units, as elaborated in \cite{Abdelhamid2015}. In the context of an incompressible XMHD plasma, the equations of motion can be effectively adapted by applying the incompressibility condition $\nabla \cdot \mathbf{v} = 0$. This adjustment simplifies the mathematical treatment and allows the equations to be recast into the following form:
\begin{equation}\label{mom}
\rho \left[ \frac{\partial \mathbf{v}}{\partial t} + (\mathbf{\nabla \times v}) \times \mathbf{v} \right] = \left( \mathbf{\nabla} \times \mathbf{B} \right) \times \mathbf{B}^{\ast} - \mathbf{\nabla} p - \rho \mathbf{\nabla} \left( \frac{v^{2}}{2} + d_{e}^{2} \frac{\left( \mathbf{\nabla} \times \mathbf{B} \right)^{2}}{2\rho^{2}} \right),
\end{equation}

\begin{equation}\label{ohm}
\frac{\partial \mathbf{B}^{\ast}}{\partial t} = \mathbf{\nabla} \times (\mathbf{v} \times \mathbf{B}^{\ast}) - d_{i} \mathbf{\nabla} \times \left[ \frac{\left( \mathbf{\nabla \times B} \right)}{\rho} \times \mathbf{B}^{\ast} \right] + d_{e}^{2} \mathbf{\nabla} \times \left[ \frac{\left( \mathbf{\nabla} \times \mathbf{B} \right)}{\rho} \times \left( \mathbf{\nabla \times v} \right) \right],
\end{equation}

\begin{equation}\label{incom}
\mathbf{\nabla} \cdot \mathbf{v} = 0, \quad \mathbf{\nabla} \cdot \mathbf{B} = 0,
\end{equation}
With, 
\begin{equation*}
\ \mathbf{B}^{\ast }=\mathbf{B}+d_{e}^{2}\mathbf{\nabla }\times \frac{\left( 
\mathbf{\nabla }\times \mathbf{B}\right) }{\rho },
\end{equation*}

the system is closed by the pressure equation 
\begin{equation} \label{pressure}
    \frac{\partial p }{\partial t} + \mathbf{v} \cdot \nabla p = 0.
\end{equation}

where $\rho$ represents the overall mass density, $\mathbf{v}$ stands for the center-of-mass velocity, $\mathbf{B}$ denotes the magnetic field, and $p$ represents the total barotropic pressure. Additionally, $d_{e}=\sqrt{\epsilon}d_{i}$ is the electron skin depth and $d_{i}$ is the ion skin depth. Here, $\epsilon=\frac{m_e}{m_i}$ represents the ratio of electron-to-ion mass. Before proceeding further, we want to point out that, despite the fact that XMHD is a much more encompassing model than traditional MHD, physical effects such as anisotropy, compressibility, dissipation, and kinetic effects (such as Landau damping) will not be incorporated into our analysis. We intend to make them the focus of our upcoming investigations. 
\section{Incompressible XMHD waves}\label{waves}

 The purpose of this section is to explore the propagation of waves in incompressible XMHD using the geometric optics approach, a mathematical framework for describing wavefronts and rays. This is done in order to analyze the behavior of these waves as they move through an incompressible plasma environment \cite{KravtsovOrlov1990, Bernstein1975, Hameiri2005}. The goal of using this method is to acquire a deeper comprehension of the dispersion characteristics and stability of various wave modes in XMHD, taking into consideration changes caused by small-scale effects.

To initiate our analysis, we consider an inhomogenous equilibrium state of the plasma and introduce small perturbations to this equilibrium. The dynamic variables can be expressed in the form:

\begin{equation}   \label{linear}
\left.
\begin{aligned}
&\mathbf{v }=\mathbf{v }_{0}+\mathbf{\tilde{v}},\\
&\mathbf{B}=\mathbf{B}_{0}+\mathbf{\tilde{B}}, \\
& p= p_{0} + \tilde{p},\\
&\rho = \rho_{0}\\
\end{aligned}
\right\}
\quad
\end{equation}
 
where subscript 0 denotes the spatio-temporary inhomogenous equilibrium fields, and the tilde represents small perturbations. 

Now, we will utilize the linear method described in \ref{linear} for the incompressible XMHD system Eqs, (\ref{mom})-(\ref{pressure}). This will result in a system of equations that the perturbations must satisfy,

\begin{eqnarray} \label{lmom}
\rho _{0}\frac{\partial \mathbf{\tilde{v}}}{\partial t}+\rho _{0}((\mathbf{\nabla }\times \mathbf{v }_{0})\times \mathbf{\tilde{v}})+\rho _{0}((\mathbf{\nabla }\times \mathbf{\tilde{v}})\times \mathbf{v}_{0})&=&(\mathbf{\nabla }\times \mathbf{B}_{0})\times \mathbf{\tilde{B}}^{\ast }+(\mathbf{\nabla }\times \mathbf{\tilde{B})\times B}_{0}^{\ast }-\mathbf{\nabla} \tilde{p}-\rho _{0}\mathbf{\nabla (v }_{0}\mathbf{\cdot \tilde{v})}\nonumber \\
&-&\rho _{0}\mathbf{\nabla}\left( d_{e}^{2}\frac{(\nabla \times \mathbf{B}_{0})(\mathbf{\nabla }\times \mathbf{\tilde{B})}}{\rho_{0}^{2}}\right), 
\end{eqnarray}

\begin{eqnarray}\label{lohm}
\frac{\partial \mathbf{\tilde{B}}^{\ast }}{\partial t}=\nabla \times (\mathbf{v}_{0}\times \mathbf{\tilde{B}}^{\ast })&+&\nabla \times (\mathbf{\tilde{v}}\times \mathbf{B}_{0}^{\ast})-d_{i}\mathbf{\nabla }\times \left[\frac{\left( \mathbf{\nabla \times B}_{0}\right) \times \mathbf{\tilde{B}}^{\ast }}{\rho _{0}}\right] -d_{i}\mathbf{\nabla }\times \left[ \frac{(\mathbf{\nabla \times \tilde{B}})\times \mathbf{B}_{0}^{\ast }}{\rho _{0}}\right]\nonumber \\  
&+&d_{e}^{2}\mathbf{\nabla }\times \left[ \frac{(\mathbf{\nabla }\times \mathbf{B}_{0})}{\rho_{0}}\times (\mathbf{\nabla \times \tilde{v}})\right] +d_{e}^{2}\mathbf{\nabla }\times \left[ \frac{(\mathbf{\nabla }\times \mathbf{\tilde{B}})}{\rho _{0}}\times (\mathbf{\nabla \times v }_{0})\right] 
\end{eqnarray}
and, 

\begin{equation}\label{lp}
    \frac{\partial \tilde{p} }{\partial t} + \mathbf{v}_{0} \cdot \nabla \tilde{p} +\mathbf{\tilde{v}}\cdot \nabla p_{0}  = 0.
\end{equation}

To implement the geometric optics ansatz, each perturbed quantity is assumed to exhibit rapid variation across magnetic field lines and is represented using the geometric optics method with the expression $e^{i\varphi / \delta}$, where $\delta$ is a small parameter such that $\delta \ll 1 $. The phase function, denoted by $\varphi = \varphi(\mathbf{x}, t)$, is common to all perturbed profiles, with $\mathbf{x}$ representing the position vector. Consequently, the perturbations in the magnetic, velocity fields and pressure, denoted by $\mathbf{\tilde{B}}$, $\mathbf{\tilde{v}}$ and, $\tilde{p}$ are expressed as a series expansion, each term modulated by this rapidly varying phase function:
\begin{equation}\label{geoptics}
\mathbf{\tilde{\psi}}=\begin{bmatrix}
\mathbf{\tilde{v}} \\
\mathbf{\tilde{B}}\\
\tilde{p}
\end{bmatrix} = e^{i\varphi \delta} \sum_{j=0}^{\infty}\mathbf{\tilde{\psi}}^{j}.
\end{equation}

The coefficients $\mathbf{\tilde{\psi}}^{j}$, where $j = 0, 1, 2, \ldots $, correspond to the amplitude of the wave at different orders of $\delta$. This formulation, as defined in Equation (\ref{geoptics}), holds under the prerequisite that the local wavelength $ 2\pi \delta / |\nabla \varphi| $ remains substantially smaller than the scale over which the equilibrium state changes. This criterion is essential to ensure that perturbations are perceived as localized disturbances against a comparatively steady background, which is crucial for accurately describing the dynamics influenced by the Hall effect and electron inertia. The assumption here is that the equilibrium state alters over a macroscopic length scale, which is significantly larger than both the ion and electron skin depths \cite{Hameiri2005}. This setup facilitates a clearer understanding of how small-scale phenomena can impact the broader dynamics within the plasma environment.

To incorporate the effects of Hall drift and electron inertia effectively, we choose a scaling such that the ion skin depth $d_i$ and the squared electron skin depth $d_e^2 $are scaled as $ d_i = \delta \bar{d}_i$ and  $d_e^2 = \delta^2 \bar{d}_e^2 $, respectively. This scaling allows these effects to emerge prominently in the leading-order dynamics of the plasma.
Now, substituting ansatz (\ref{geoptics}) in the linearized incompressible XMHD equations (\ref{lmom})-(\ref{lp}), yields the following result.
\begin{eqnarray}
\rho _{0}\frac{\partial \mathbf{\tilde{v}}^{0}}{\partial t}&+&\frac{i}{\delta }\varphi_{t} \rho_{0} \mathbf{\tilde{v}}^{0} + \rho _{0}((\mathbf{\nabla } \times \mathbf{v}_{0}) \times \mathbf{\tilde{v}}^{0}) + \rho _{0}((\mathbf{\nabla } \times \mathbf{\tilde{v}}^{0}) \times \mathbf{v}_{0})+\frac{i}{\delta }\rho _{0}((\mathbf{\nabla }\varphi \times \mathbf{\tilde{v}}^{0})\times \mathbf{v}_{0})=\nonumber\\ &&(\mathbf{\nabla }\times \mathbf{B}_{0})\times \mathbf{\tilde{B}}^{\ast 0}+(\mathbf{\nabla }\times\mathbf{\tilde{B}}^{0}\mathbf{)\times B}_{0}^{\ast }+\frac{i}{\delta }(\mathbf{\nabla }\varphi \times \mathbf{\tilde{B}}^{0}\mathbf{)\times B}_{0}^{\ast }-\frac{i}{\delta } \mathbf{\nabla}\varphi \tilde{p} -\rho _{0}\mathbf{\nabla (v}_{0}\mathbf{\cdot \tilde{v}}^{0}\mathbf{)} \nonumber \\ && -\frac{i}{\delta }\rho _{0}\mathbf{(v}_{0}\cdot \mathbf{\tilde{v}}^{0}\mathbf{)\nabla }\varphi -i\delta \bar{d}_{e}^{2}\left( \frac{(\nabla \times \mathbf{B}_{0})(\mathbf{\nabla }\times \mathbf{\tilde{B}}^{0})}{\rho _{0}}\right) \mathbf{\nabla }\varphi -\delta ^{2}\bar{d}_{e}^{2}\rho _{0}\mathbf{\nabla }\left( \frac{(\nabla \times \mathbf{B}_{0})(\mathbf{\nabla }\times \mathbf{\tilde{B}}^{0})}{\rho_{0}^{2}}\right)\nonumber \\ && +\bar{d}_{e}^{2}\left( \frac{(\nabla \times \mathbf{B}_{0})(\mathbf{\nabla }\varphi \times\mathbf{\tilde{B}}^{0})}{\rho _{0}}\right) \mathbf{\nabla }\varphi +i\delta \bar{d}_{e}^{2}\rho _{0}\mathbf{\nabla }\left( \frac{(\nabla \times \mathbf{B}_{0})(\mathbf{\nabla }\varphi \times \mathbf{\tilde{B})}}{\rho _{0}^{2}}\right) ,
\end{eqnarray}

\begin{eqnarray}
\frac{\partial \mathbf{\tilde{B}}^{\ast 0}}{\partial t}&&+\frac{i}{\delta }\varphi _{t}\mathbf{\tilde{B}}^{\ast 0} =\nabla \times (\mathbf{v }_{0}\times \mathbf{\tilde{B}}^{\ast 0})+\frac{i}{\delta }\mathbf{\nabla }\varphi \times (\mathbf{v }_{0}\times \mathbf{\tilde{B}}^{\ast 0})+\nabla\times (\mathbf{\tilde{v}}^{0}\times \mathbf{B}_{0}^{\ast })+\frac{i}{\delta }\mathbf{\nabla }\varphi \times (\mathbf{\tilde{v}}^{0}\times \mathbf{B}_{0}^{\ast })  \nonumber \\ &&-\delta \bar{d}_{i}\mathbf{\nabla }\times \left[ \frac{\left(\mathbf{\nabla \times B}_{0}\right) \times \mathbf{\tilde{B}}^{\ast 0}}{\rho_{0}}\right] -i\bar{d}_{i}\mathbf{\nabla }\varphi \times \left[ \frac{\left( \mathbf{\nabla \times B}_{0}\right) \times \mathbf{\tilde{B}}^{\ast 0}}{\rho _{0}}\right]-\delta \bar{d}_{i}\mathbf{\nabla }\times \left[ \frac{(\mathbf{\nabla \times \tilde{B}}^{0})\times \mathbf{B}_{0}^{\ast }}{\rho _{0}}\right] \nonumber \\ &&-i\bar{d}_{i}\mathbf{\nabla }\varphi \times \left[ \frac{(\mathbf{\nabla \times \tilde{B}}^{0})\times \mathbf{B}_{0}^{\ast }}{\rho_{0}}\right] -i\bar{d}_{i}\mathbf{\nabla }\times \left[\frac{(\mathbf{\nabla }\varphi \mathbf{\times \tilde{B}}^{0})\times \mathbf{B}_{0}^{\ast }}{\rho_{0}}\right] +\frac{\bar{d}_{i}}{\delta }\mathbf{\nabla }\varphi \times\left[ \frac{(\mathbf{\nabla }\varphi \mathbf{\times \tilde{B}}^{0})\times \mathbf{B}_{0}^{\ast }}{\rho_{0}}\right] \nonumber \\ && +\delta^{2} \bar{d}_{e}^{2}\mathbf{\nabla }\times \left[\frac{(\mathbf{\nabla }\times \mathbf{B}_{0})}{\rho _{0}}\times (\mathbf{\nabla \times \tilde{\mathbf{v}}}^{0})\right] +i\delta\bar{d}_{e}^{2}\mathbf{\nabla }\varphi \times \left[\frac{(\mathbf{\nabla }\times \mathbf{B}_{0})}{\rho_{0}}\times (\mathbf{\nabla \times\tilde{\mathbf{v}}}^{0})\right] \nonumber \\ &&+i\delta\bar{d}_{e}^{2}\mathbf{\nabla }\times \left[ \frac{(\mathbf{\nabla }\times\mathbf{B}_{0})}{\rho _{0}}\times (\mathbf{\nabla}\varphi\mathbf{\times\tilde{\mathbf{v}}}^{0})\right]+i\bar{d}_{e}^{2}\mathbf{\nabla }\varphi \times \left[ \frac{(\mathbf{\nabla }\times \mathbf{B}_{0})}{\rho_{0}}\times (\mathbf{\nabla }\varphi \mathbf{\times \tilde{\mathbf{v}}}^{0})\right]\nonumber \\ && +\delta^{2}\bar{d}_{e}^{2}\mathbf{\nabla }\times \left[\frac{(\mathbf{\nabla }\times \mathbf{\tilde{B}}^{0})}{\rho _{0}}\times (\mathbf{\nabla \times \nu }_{0})\right]  +i\delta \bar{d}_{e}^{2}\mathbf{\nabla }\varphi \times \left[\frac{(\mathbf{\nabla }\times \mathbf{\tilde{B}}^{0})}{\rho _{0}}\times (\mathbf{\nabla \times \mathbf{v} }_{0})\right]\nonumber \\ && +i\delta \bar{d}_{e}^{2}\mathbf{\nabla }\times \left[ \frac{(\mathbf{\nabla }\varphi \times \mathbf{\tilde{B}}^{0})}{\rho _{0}}\times (\mathbf{\nabla \times \mathbf{v} }_{0})\right] +i\bar{d}_{e}^{2}\mathbf{\nabla }\varphi \times \left[ \frac{(\mathbf{\nabla }\varphi \times \mathbf{\tilde{B}}^{0})}{\rho _{0}}\times (\mathbf{\nabla \times \mathbf{v}}_{0})\right] ,
\end{eqnarray}
where, at $O(1)$%
\begin{equation*}
\mathbf{\tilde{B}}^{\ast0}=\mathbf{\tilde{B}}^{0}+\bar{d}_{e}^{2}\mathbf{\nabla }\varphi \times \frac{\left( \mathbf{\nabla }\varphi \times \mathbf{\tilde{B}}^{0}\right) }{\rho _{0}},
\end{equation*}
and, 

\begin{equation}\label{lp0}
    \frac{\partial \tilde{p}^{0} }{\partial t}+ \frac{i}{\delta }\varphi_{t} \tilde{p}^{0}+ \frac{i}{\delta }(\mathbf{v}_{0} \cdot \nabla\varphi) \tilde{p}^{0} +\mathbf{\tilde{v}}\cdot \nabla p_{0}  = 0.
\end{equation}

By equating the orders of $\delta $, we find that at $O(1/\delta )$, 

\begin{equation}\label{order1}
-(\mathbf{k}\times \mathbf{\tilde{B}}^{0})\times \mathbf{B}_{0}-\omega \rho_{0}\mathbf{\tilde{v}}^{0}=0,  
\end{equation}

\begin{equation}\label{order2}
\rho _{0}^{-1}\bar{d}_{e}^{2}\omega \mathbf{k}\times (\mathbf{k}\times \mathbf{\tilde{B}}^{0})+i\rho _{0}^{-1}\bar{d}_{i}(\mathbf{k}\cdot \mathbf{B}_{0})\mathbf{k}\times\mathbf{\tilde{B}}^{0}-\omega \mathbf{\tilde{B}}^{0}-(\mathbf{k}\cdot \mathbf{B}_{0})\mathbf{\tilde{v}}^{0}=0 ,
\end{equation}
with,

\begin{equation}\label{lp01}
   - \omega\tilde{p}^{0}= 0.
\end{equation}

In the described equations, the notation $-(\varphi_ {t}+\mathbf{v}_{0}\cdot \nabla \varphi )=\omega$ and $\nabla\varphi =\mathbf{k}$ is employed to symbolize the wave frequency and the local wave vector, respectively. These notations are critical for defining the traits of wave propagation within the plasma. As detailed in \cite{Hameiri2005}, $\omega (\mathbf{x},t)$ denotes the local wave frequency, which varies with both position and time, effectively capturing the temporal variations of the wave within the plasma. Similarly, $\mathbf{k}(\mathbf{x},t)$ defines the local wave vector, providing insights into the direction and magnitude of wave propagation at every point in space and time. It is important to note that equation (\ref{lp01}) suggests the only feasible solution is $\tilde{p} = 0$, indicating that  the pressure equation does not influence the dispersion relation in XMHD under incompressible conditions, as pressure perturbations are effectively separated from wave dynamics in this limit. Consequently, the remaining system Eqs. (\ref{order1})--(\ref{order2}) can be reformulated as a Hermitian matrix eigenvalue problem for $\omega$ (for a given $\mathbf{k}$), expressed as:
\begin{equation*}
A \mathbf{u} = 0,
\end{equation*}
where $\mathbf{u}$ is defined as the transpose vector $(\mathbf{\tilde{v}}^{0}, \mathbf{\tilde{B}}^{0})^{T}$, representing the zeroth-order perturbations in velocity and magnetic field. This matrix representation permits a systematic method to solve for $\omega$, aiding in the analysis of wave behavior in the plasma by determining the system's eigenvalues and eigenvectors. The Hermitian matrix $A$ is structured as follows:
\begin{equation}\label{HA}
A=
\begin{bmatrix}
-\omega \rho _{0}I & \mathbf{kB}_{0}^{T}-(\mathbf{k\cdot B}_{0})I \\ \mathbf{B}_ {0}\mathbf{k}^{T}-(\mathbf{k\cdot B}_{0})I & -\omega I-\rho_{0}^{-1}\bar{d}_{e}^{2}\omega k^{2}I+i\rho_{0}^{-1}\bar{d}_{i}(\mathbf{k\cdot B}_{0})\mathbf{k\times }.
\end{bmatrix}
\end{equation}

In this context, $k=\left \vert \mathbf{k}\right \vert $ indicates the magnitude of the local wave vector $\mathbf{k}$. The superscript T denotes the transpose of a vector such that $\mathbf{k}^{T} \mathbf{u}=\mathbf{k}\cdot \mathbf{u}$ represents the product of the dots of $\mathbf{k}$ with the vector $\mathbf{u}$. Additionally, $I$ refers to a $3\times3$ identity matrix, featuring 1s on the diagonal and 0s elsewhere. It should also be noted that $A$ is a matrix $6 \times 6$, with each of the four entries being matrices themselves. The equation $A\mathbf{u} = 0$ can be viewed as an eigenvalue problem for $\omega$, formulated as $S\mathbf{u} = \omega P\mathbf{u}$, where $S$ is Hermitian, and $P$ is symmetric and positive definite.While the cross product operation $\mathbf{k} \times $ in the diagonal term of $S$ is inherently anti-Hermitian, introducing the imaginary unit $i$ brings it in line with the characteristics of a Hermitian matrix. Consequently, for any real value of $k$, there are six corresponding real $\omega$ values, reflecting the scenarios found in both magnetohydrodynamics (MHD) and Hall MHD \cite{Hameiri2005,goedbloed2004principles}. This shows the presence of six specific wave solutions, in accordance with the structure represented in equation (\ref{HA}), and their individual frequencies are determined by the solutions $\omega$ found in the dispersion relation $\det A = 0$. The eigenvector $\mathbf{u}$ connected to each $\omega$ communicates the details of the wave, thereby characterizing its propagation and effects.To solve the equation $A\mathbf{u} = 0$ and derive the wave solutions, various linear operations are performed on the rows of matrix $A$. The detailed procedures to reformulate and solve equation (\ref{HA}) further enhance the comprehension and analysis of wave dynamics within the system. Equation (\ref{HA}) can be rewritten as

\begin{equation}\label{A}
A=
\begin{bmatrix}
Q & R \\ 
R^{T} & S
\end{bmatrix}
\end{equation}
Where $Q$, \ $R$ and, $S$ \ are $3\times 3$\ matrices, which are defined by,
\begin{eqnarray*}
    Q&=&-\omega \rho _{0} I\\
    R&=&\mathbf{kB}_{0}^{T}-(\mathbf{k\cdot B}_{0})I\\  
    R^{T}&=&\mathbf{B}_{0}\mathbf{k}^{T}-(\mathbf{k\cdot B}_{0})I\\
    S&=&-\omega I-\rho _{0}^{-1}\bar{d}_{e}^{2}\omega k^{2}I+i\rho _{0}^{-1}\bar{d}_{i}(\mathbf{k\cdot B}_{0})\mathbf{k\times}
\end{eqnarray*}

By incorporating $(-R^{T}Q^{-1})$ times the first row into the second row in equation (\ref{A}), we obtain the decoupled equation $\acute{S}\tilde{B}^{0} = 0$, where
\begin{equation}
\acute{S} = S - R^{T}Q^{-1}R.
\end{equation}
This manipulation simplifies the system, allowing for a more straightforward analysis of the dynamics. Consequently, equation (\ref{A}) can now be rewritten in a more tractable form, clarifying the relationships and dependencies among the variables:
\begin{eqnarray}\label{A2}
\begin{bmatrix}
Q & R \\
0 & \acute{S}
\end{bmatrix}
\begin{bmatrix}
\mathbf{\tilde{v}}^{0} \\
\mathbf{\tilde{B}}^{0}
\end{bmatrix}
= 0.
\end{eqnarray}

 Similarly, the relationship
$Q\mathbf{\tilde{v}}^{0} + R\tilde{\mathbf{B}}^{0} = 0$
is established with the solution given by

\begin{equation} \label{v-vector}
\mathbf{\tilde{v}}^{0} = -Q^{-1}R\tilde{\mathbf{B}}^{0},
\end{equation}

which can be determined once $\tilde{\mathbf{B}}^{0}$ is specified. Following this, the determinant of the matrix $A$ in equation (\ref{A2}) can be expressed as a product of the determinants of the matrices involved, specifically, 
\begin{equation}\label{detA}
    \det A = \det Q \cdot \det \acute{S}.
\end{equation}

The determinant $\det Q$ is calculated as the product of its three eigenvalues $\lambda_{j}$ with $j = 1, 2, 3$. These eigenvalues are derived from the eigenvalue equation
$Q\mathbf{\tilde{v}}^{0} = \lambda_{j}\mathbf{\tilde{v}}^{0},$
where $Q = -\omega \rho_{0} I$. One of the eigenvectors of $Q$, denoted by $\mathbf{\tilde{v}}^{0}_{1}$, is $\mathbf{k}$, and the remaining two eigenvectors are orthogonal to $\mathbf{k}$, with all three eigenvalues given by 
$\lambda_{1} = \lambda_{2} = \lambda_{3} = -\omega \rho_{0}.$
Then,
\begin{equation} \label{detQ}
\det Q=\lambda _{1}\cdot \lambda _{2}\cdot \lambda _{3}=-\rho _{0}^{3}\omega ^{3}
\end{equation}

To calculate the value of $\det \acute{S}$, which arises from the product of three eigenvalues in the eigenvector equation $\acute{S}\tilde{\mathbf{B}}^{0} = \lambda_{j}\tilde{\mathbf{B}}^{0}$, we consider the composition of the perturbed magnetic field $\tilde{\mathbf{B}}^{0}$. This field includes three components: one aligned parallel to the propagation vector $\mathbf{k}$ and the other two orthogonal to it. Taking the scalar product of the eigenvector equation with the propagation vector $\mathbf{k}$ results in:

\begin{equation}\label{sii}
\acute{S}(\mathbf{k} \cdot \tilde{\mathbf{B}}^{0}) = -\omega (\mathbf{k} \cdot \tilde{\mathbf{B}}^{0}) - \rho_{0}^{-1} \bar{d}_{e}^{2} \omega k^{2} (\mathbf{k} \cdot \tilde{\mathbf{B}}^{0})  
\end{equation}

From this we deduce that $\lambda_{1} = -\omega^{\ast}$, where $\omega^{\ast} = \omega (1 + \rho_{0}^{-1} \bar{d}_{e}^{2} k^{2})$ represents a modified frequency factor that accounts for the influence of electron inertia.

Similarly, to get the two eigenvalues in the perpendicular direction to the propagation vector $\mathbf{k}$, we take the vector product of the eigenvector equation with the propagation vector $\mathbf{k}$, this operation yields the following.

\begin{equation}\label{sp}
\acute{S}_{P}(\mathbf{k\times }\tilde{\mathbf{B}}^{0})=-\omega (\mathbf{k\times \tilde{B}}^{0})+i\rho _{0}^{-1} \bar{d}_{i}\mathbf{(k\cdot B}_{0}\mathbf{)k\times \tilde{B}}^{0}-\rho _{0}^{-1}\bar{d}_{e}^{2}\omega k^{2}(\mathbf{k\times \tilde{B}}^{0})+\frac{1}{\omega \rho _{o}}(\mathbf{k\cdot B}_{0})^{2}( \mathbf{k\times \tilde{B}}^{0})  
\end{equation}

In this context, we establish $\mathbf{B}_{P}$ as the part of the equilibrium field $\mathbf{B}_{0}$ that is perpendicular to $\mathbf{k}$, denoted as $\mathbf{B}_{P}\equiv \mathbf{B}_{0}-\frac{1}{k^{2}}(\mathbf{k}\cdot \mathbf{B}_{0})^{2}\mathbf{k}$. We will also use the vectors $\mathbf{B}_{P}$ and $\mathbf{k}\times \mathbf{B}_{0}/k$ as the foundation vectors for this subspace, so that each vector in the subspace can be expressed as

\begin{equation}
 \mathbf{\tilde{B}}^{0}=x\mathbf{B}_{P}+y\mathbf{k}\times \mathbf{B_{0}}/k
\end{equation}
 as an operator operating in $x-y$ space, equation (\ref{sp}) can be written as

\begin{eqnarray}
\acute{S}_{p}(\mathbf{k\times }( x\mathbf{B}_{P}+y\mathbf{k}\times \mathbf{B}/k ) &=&-\omega (\mathbf{k\times (}x\mathbf{B}_{P}+y\mathbf{k}\times \mathbf{B}/k)+i\rho _{0}^{-1}\bar{d}_{i}\mathbf{(k\cdot B}_{0}\mathbf{)[k\times (}x\mathbf{B}_{P}+y\mathbf{k}\times \mathbf{B}/k)] \nonumber \\
&&-\rho _{0}^{-1}\bar{d}_{e}^{2}\omega k^{2}(\mathbf{k\times (}x\mathbf{B}_{P}+y\mathbf{k}\times \mathbf{B}/k)\nonumber \\
&&+\frac{1}{\omega \rho _{o}}(\mathbf{k\cdot B}_{0})^{2}[\mathbf{k\times (}x\mathbf{B}_{P}+y\mathbf{k}\times \mathbf{B}/k)].
\end{eqnarray}

Then

\begin{eqnarray} \label{Sp}
\acute{S}_{p}[(\mathbf{k\times B}_{P})x+(\mathbf{k\times k}\times \mathbf{B}/k)y] &=&-\omega \lbrack (\mathbf{k\times B}_{P})x+(\mathbf{k\times k}\times \mathbf{B}/k)y] + i\rho_{0}^{-1} \bar{d}_{i} \mathbf{(k\cdot B}_{0}\mathbf{)}[(\mathbf{k\times B}_{P})x  \nonumber \\
&&+(\mathbf{k\times k}\times \mathbf{B}/k)y]-\rho _{0}^{-1}\bar{d}_{e}^{2}\omega k^{2}[(\mathbf{k\times B}_{P})x+(\mathbf{k\times k}\times \mathbf{B}/k)y] \nonumber \\
&&+\frac{1}{\omega \rho _{o}}(\mathbf{k\cdot B}_{0})^{2}[(\mathbf{k\times B}_{P})x+(\mathbf{k\times k}\times \mathbf{B}/k)y]
\end{eqnarray}

After some algebraic manipulations, equation (\ref{Sp}) can be represented in matrix form

\begin{equation}\label{spmatrix}
\acute{S}_{p}=
\begin{bmatrix}
\acute{S}_{11} & -i\rho _{0}^{-1}\bar{d}_{i}\mathbf{k(k\cdot B}_{0}\mathbf{)}
\\ 
i\rho _{0}^{-1}\bar{d}_{i}\mathbf{k(k\cdot B}_{0}\mathbf{)} & \frac{1}{\omega \rho _{o}}[(\mathbf{k\cdot B}_{0})^{2}-\rho _{0}\omega \omega^{\ast}]
\end{bmatrix},
\end{equation}

wheres $\acute{S}_{11}=\frac{1}{\omega \rho _{o}k^{2}}[k^{2}(\mathbf{k\cdot B}_{0})^{2}-\rho _{0}\omega \omega ^{\ast}k^{2}]$.

Now, we can evaluate $\det A$ based on equation (\ref{detA}), where we observe that $\det \acute{S} = - \omega^{\ast} \det \acute{S}_{p}$. Recalling equation (\ref{detQ}), with the aid $\det\acute{S}_{p}$ derived from the matrix (\ref{spmatrix}), we can determine $\det A$ as follows:

\begin{equation}
\det A=
(\omega^{3}\rho _{0}^{3})\left( \frac{\omega^{\ast}}{\omega^{2}\rho _{0}^{2}}\left( (\mathbf{k}\cdot \mathbf{B}_{0}^{2})-\rho _{0}\omega \omega ^{\ast }\right)^{2}-\frac{\bar{d}_{i}^{2} \omega^{\ast}}{\rho _{0}^{2}}k^{2}\mathbf{(k\cdot B}_{0}\mathbf{)}^{2}\right),
\end{equation}.

Then, after performing a series of straightforward manipulations, the dispersion relation for incompressible XMHD can be expressed in the following form:

\begin{equation}\label{dispersion}
\left((1+\frac{\bar{d}_{e}^{2}}{\rho_{0}} k^{2}) \rho _{0}\omega ^{2}-(\mathbf{k}\cdot \mathbf{B}_{0})^{2}\right)^{2}=\bar{d}_{i}^{2}\omega^{2}k^{2}\mathbf{(k}\cdot \mathbf{B}_{0})^{2} 
\end{equation}

To determine the eigenvectors corresponding to $A\mathbf{u}=0$ with $\omega \neq 0$, then

\begin{equation}\label{eigenvector}
\mathbf{\tilde{B}}^{0}=\alpha \left[ \mathbf{B}_{p} + \frac{i\bar{d}_{i}\omega (\mathbf{k}\cdot \mathbf{B}_{0})}{\rho _{0}\omega \omega^{\ast}-(\mathbf{k\cdot B}_{0})^{2}}\mathbf{k} \times \mathbf{B}_{0}\right]   
\end{equation}

where $\alpha = \frac{\omega^{2}(\bar{d}_{e}^{2}k^{2}-\rho _{0})-(\mathbf{k\cdot B}_{0})^{2}}{\rho _{0}\omega }$, from which the eigenvector $\mathbf{\tilde{v}}^{0}$ can be determined using the equation $\mathbf{\tilde{v}}^{0}=- Q^{-1}R\mathbf{\tilde{B}}^{0}$.

\subsection{Limits of Incompressible XMHD}
Now, we will examine the different regimes of interest for the dispersion relations (\ref{dispersion}).

\subsubsection{Hall MHD Regime:}
Let us consider the scenario where Hall effects are significant but electron inertia is negligible, $d_{e} \rightarrow 0$, known as the Hall regime. Under this condition, the dispersion (\ref{dispersion}) simplifies to,
\begin{equation} \label{Hdis}
\left( \rho_{0}\omega^{2}-(\mathbf{k}\cdot \mathbf{B}_{0})^{2}\right)^{2}-\bar{d}_{i}^{2}\omega^{2} k^{2}(\mathbf{k}\cdot \mathbf{B}_{0})^{2}=0,
\end{equation}
indicating two branches: the shear-Whistler mode and the magnetosonic-cyclotron mode. This dispersion relation matches the nonlinear Hall MHD dispersion derived in earlier works \cite{Abdelhamid2016, Mahajan2005, Sahraoui2007} under constant background fields.

\subsubsection{Ideal MHD Regime:}
Consider the limits where both electron inertia and Hall effect are negligible, i.e., $d_{i} \rightarrow 0$, $d_{e} \rightarrow 0$. This signifies operation in the ideal MHD domain. Thus, the dispersion relation for ideal MHD can be obtained as,
\begin{equation}\label{Idis}
\left( \rho_{0}\omega^{2}-(\mathbf{k}\cdot \mathbf{B}_{0})^{2}\right)^{2}=0,
\end{equation}
which corresponds to the shear Alfvén waves of ideal MHD, as noted in \cite{goedbloed2004principles}.
\begin{figure}[ht!]
\centering
\includegraphics[width=10.5cm]{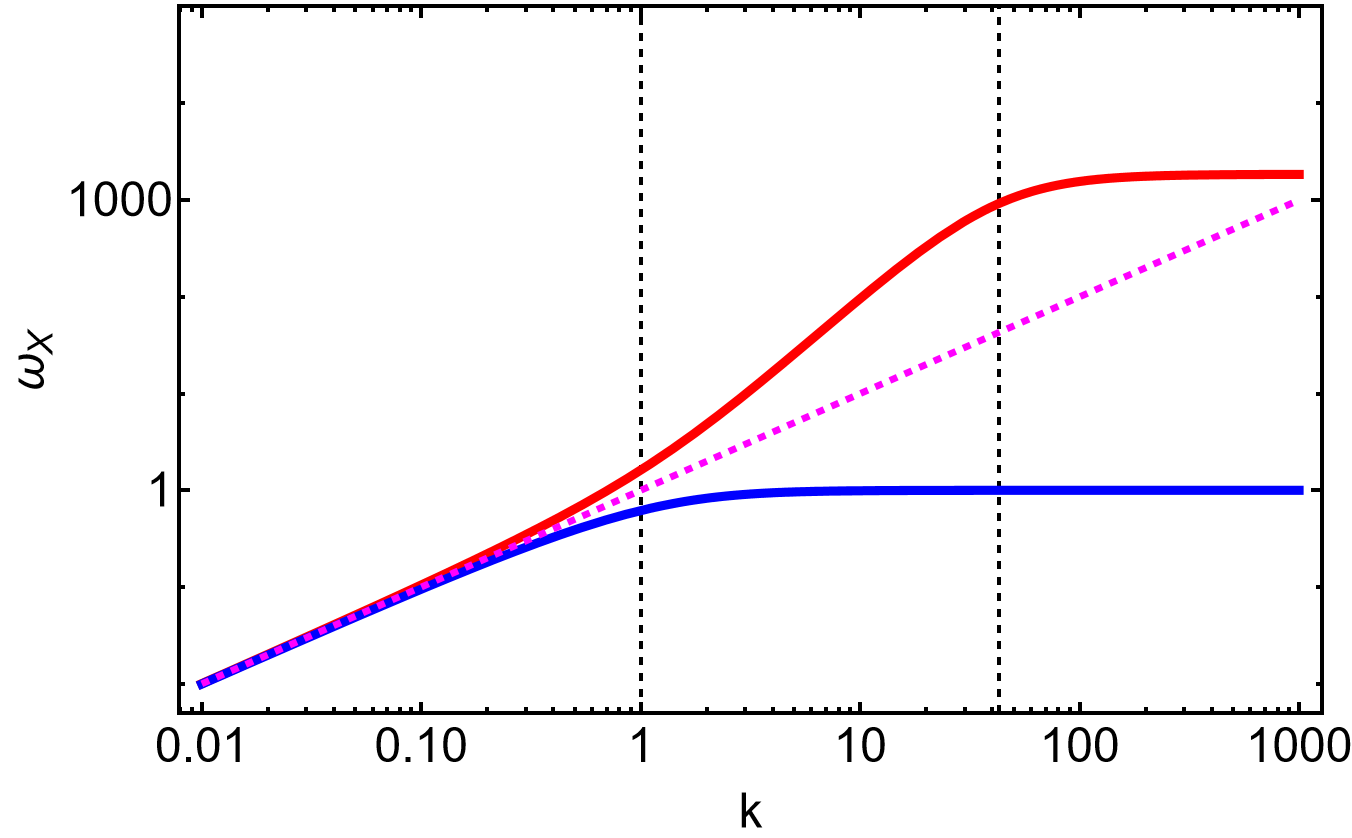}
\caption{(Color online) Dispersion relation ($\omega_X$) curves for XMHD equation (\ref{dispersion}), with parameters $d_{i}=1$, $d_{e}=0.0233$, assuming a constant equilibrium density $\rho_0 = 1$ and a wave vector $\mathbf{k}$ aligned with the magnetic field $\mathbf{B}_{0}$ such that $\mathbf{k} \cdot \mathbf{B}_{0} = k$. The upper and lower branches represent whistler and ion cyclotron waves, respectively. The dashed line represents ideal Alfvén waves, given by the solution of (\ref{Idis}). Vertical dotted lines delineate the ideal, Hall, and electron inertia regions, in left-to-right order.}
 \label{fig1}
\end{figure}
\begin{figure}[ht!]
\centering
\includegraphics[width=10.5cm]{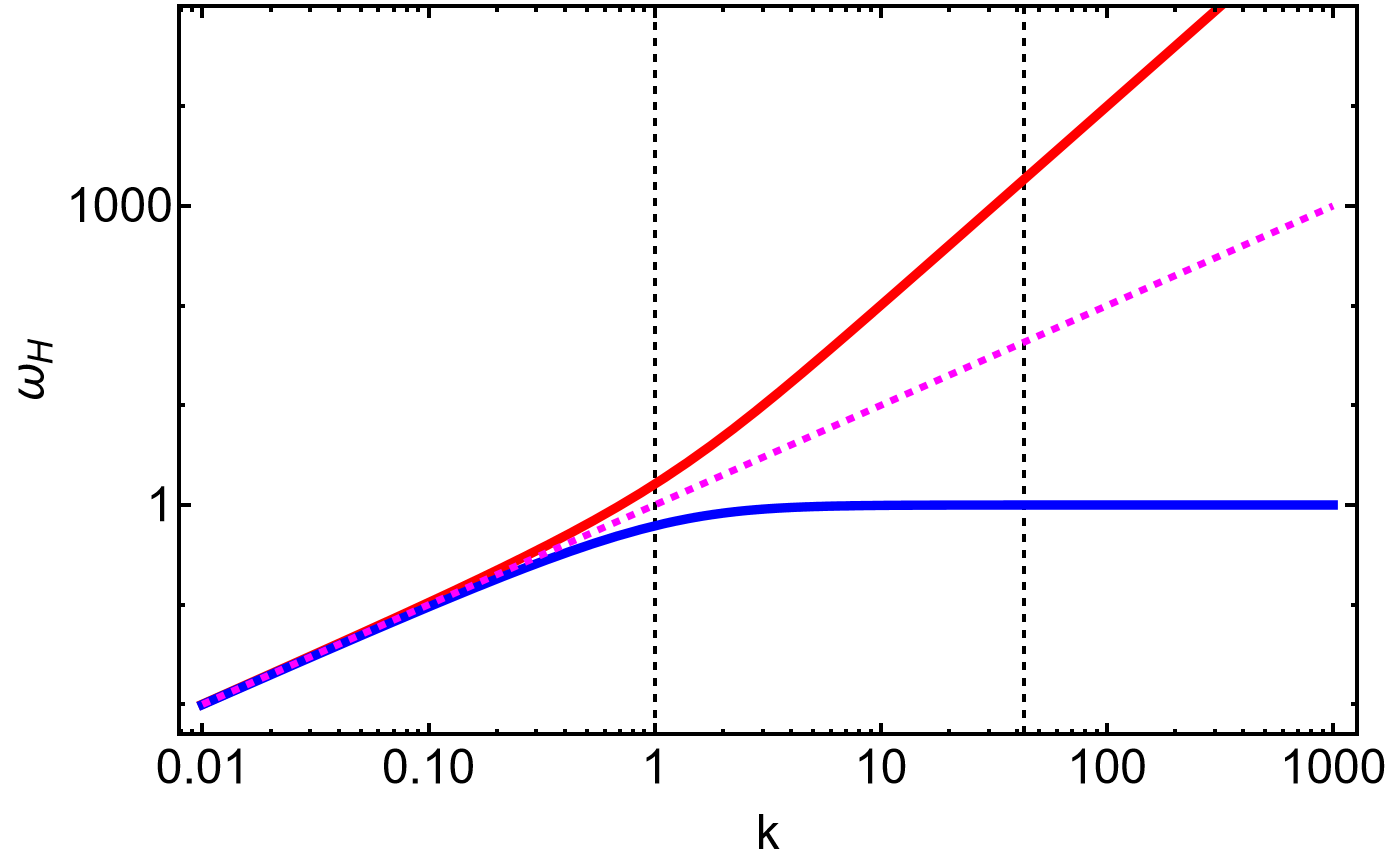}
\caption{(Color online) Dispersion relation ($\omega_H$) curves for HMHD equation (\ref{Hdis}) for $d_{i}=1$, assuming a constant equilibrium density $\rho_0 = 1$ and a wave vector $\mathbf{k}$ aligned with the magnetic field $\mathbf{B}_{0}$ such that $\mathbf{k} \cdot \mathbf{B}_{0} = k$. The upper and lower branches represent whistler and ion cyclotron waves, respectively. The dashed line represents ideal Alfvén waves, given by the solution of (\ref{Idis}). Vertical dotted lines delineate the ideal, Hall, and electron inertia regions, in left-to-right order.}
 \label{fig2}
\end{figure}

The dispersion relation (\ref{dispersion}) aligns with the nonlinear XMHD dispersion relation \cite{Abdelhamid20162}, demonstrating how electron inertia inclusion not only alters wave modes but also eliminates Hall MHD solution singularities and encapsulates more two-fluid model physics \cite{Zhao2015, Kakuwa2017, Jonghe2020, Choi2023}. The chief differences from previous works stem from our derivation of the dispersion relation and the associated wave eigenvectors within the XMHD, specially by considering an inhomogeneous equilibrium. This result extends beyond the earlier linear and nonlinear works that were limited constant equilibrium states. This put our analysis as a reasonable candidate plasma environment where the   inhomogeneities of the equilibrium fields are dominant, see for example \cite{Franci2022, Goossens2019,Kim2022, Lu2020, Lu2019, Okamoto2015, Pascoe2011, Pucci2017, Skirvin2022, Soler2015,Tzeferacos2018}. Another primary outcome of our work is the remarkable relation between the magnetic and kinetic fluctuations, as represented in equation (\ref{v-vector}). The latter relation is essential in effectively “eliminating” nonlinearity in the governing equations, allowing the linear solutions to coincide identically with the nonlinear ones \cite{Abdelhamid2016, Abdelhamid20162,Mahajan2005,Mahajan2009}. This positions our finding as a strong candidate for the investigation of processes in real plasma environments, such as the turbulence spectrum of the solar wind \cite{Magyar2019, Hahm2024}.

Figure. \ref{fig1} depicts the ion cyclotron waves in the lower branch, primarily influenced by ion dynamics and saturating at the ion gyrofrequency. This saturation indicates a maximum wave frequency determined by ion cyclotron motion, beyond which k increases yield minimal frequency changes. The upper branch represents whistler waves, propagating at higher frequencies and primarily affected by electron dynamics, saturating at the electron gyrofrequency due to electron cyclotron motion limitations. Whistler waves, a most probable electromagnetic mode observed in laboratory plasma and space \cite{Gary1993, Gurnett2017}, propagate between the electron cyclotron frequency and the lower hybrid frequency along an ambient magnetic field. Our analysis confirms that the whistler mode experiences saturation as it approaches the electron cyclotron frequency, which prevents the frequency from increasing arbitrarily. This limitation is essential for whistler modes to maintain their physical realism. In contrast, in the HMHD framework (figure. \ref{fig2}), this saturation is absent, resulting in unphysical predictions that the whistler frequency could increase indefinitely.

\section{Conclusion and future work}\label{conclusion}
Our research within the Extended Magnetohydrodynamics (XMHD) framework marks a significant advancement in understanding plasma-wave dynamics, emphasizing the integral role of small-scale physical phenomena in both astrophysical and laboratory plasma environments. By incorporating Hall drift and electron inertia effects, we have derived a nonlinear dispersion relation for incompressible XMHD and associated eigenvector solutions using the geometric optics ansatz. This achievement has shed light on the nuanced behaviors of plasma waves under XMHD conditions, revealing that electron inertia modifies the Hall MHD dispersion relation without introducing new wave branches. The dispersion relation and eigenvector solutions offer valuable insights into wave propagation and structure, potentially aiding in the interpretation of observational data from space plasma and laboratory experiments.
\\

Our study contributes a pivotal perspective to the field of plasma physics, bridging the gap between large-scale and small-scale physical phenomena. The comparison between Hall MHD and XMHD dispersion relations demonstrates that XMHD provides a more comprehensive and accurate representation of plasma dynamics, especially at higher wave numbers. Future work will extend this research to explore compressibility effects within the XMHD model, promising even more comprehensive insights into plasma wave behaviors. This research not only enriches our current understanding but also lays the groundwork for future investigations that will further unravel the complexities of plasma behavior in various contexts, from astrophysical settings to laboratory experiments.
\begin{appendices}
\section{List of Symbols}
\begin{tabular}{ll}
    $\rho$ & overall mass density\\
    $\mathbf{v}$ &  center of mass velocity\\
    $\mathbf{B}$ & magnetic field\\
    $p$ & total barotropic pressure, $p= p_{i}+ p_{e}$ \\
    $d_i$ & normalized ion skin depth \\
    $d_e$ & normalized electron skin depth, $d_{e}=\sqrt{\epsilon}d_{i}$ \\
    $\epsilon$ & electron-to-ion mass ratio, $\epsilon= \frac{m_e}{m_i}$\\
    $\mathbf{v}_0$ & inhomogeneous equilibrium velocity \\
    $\mathbf{B}_0$ & inhomogeneous equilibrium magnetic field \\
    $\tilde{}$ & tilde represents small perturbations \\
    $\mathbf{\tilde{\psi}}^{j}$ &  amplitude of the wave at different orders of $\delta$, such as $\mathbf{\tilde{B}}^{j}=\mathbf{\tilde{B}}^{0}+ \delta\mathbf{\tilde{B}}^{1}+\delta^2 \mathbf{\tilde{B}}^{2}+ \ldots$ \\
    $\delta$ & small parameter, $\delta \ll 1$ \\
    $\omega(\mathbf{x},t)$ & local wave frequency \\
    $\mathbf{k} (\mathbf{x},t)$ & local wave vector \\
    $I$ & $3\times3$ identity matrix \\
    $\mathbf{B}_{p}$ & part of equilibrium $\mathbf{B}_{0}$ perpendicular to $\mathbf{k}$ \\
    $\varphi$ & phase function \\
\end{tabular}
\end{appendices}
\bibliography{apssamp}

\end{document}